\documentclass[prl]{revtex4}
\usepackage{graphicx}

\begin{document}

\title{Pumping in a mesoscopic ring with Ahronov-Casher effect}
\author{R. Citro and F. Romeo}
\affiliation{Dipartimento di Fisica ``E. R. Caianiello'' and
Unit{\`a} C.N.I.S.M. di Salerno\\ Universit{\`a} degli Studi di
Salerno, Via S. Allende, I-84081 Baronissi (Sa), Italy}
\date{\today}

\begin{abstract}
 We investigate parametric pumping of
spin and charge currents in a mesoscopic ring interrupted by a
tunnel barrier in presence of Aharonov-Casher (AC) effect and
Aharonov-Bohm (AB) flux along the axis of the same ring.
Generation of a dc current is achieved by tuning the tunnel
barrier strength and modulating in time either a
radial(transverse) electric field or the magnetic flux. A pure
spin current is generated by the interplay of breaking spin
reversal symmetry, due to AC effect, and time-reversal symmetry
breaking, intrinsic in parametric pumping procedure. We analyze
the conditions for operating the AB-AC ring as a pure spin pump
useful in spintronics and discuss generalization of our results to
Rashba-gate-controlled rings.
\end{abstract}


\maketitle

 The physics of pumping has
attracted considerable interest in the last two decades.
Parametric pumping of electrons in mesoscopic systems refers to
the generation of a dc current by periodic modulations of two or
more system parameters (e.g., a gate voltage or a magnetic field)
in the absence of a bias voltage.  In his original work
Thouless\cite{thouless_or,niu} studied the integrated particle
current on a finite torus produced by a slow variation of the
potential and showed that the integral of the current over a
period can vary continuously, but must have an integer value in a
clean infinite periodic system with full bands.
Since then, interest in this
phenomenon has shifted to
theoretical\cite{brouwer,aleiner1,polianski,zhou} and
experimental\cite{switches,altsh} investigations of adiabatic
pumping through open quantum dots where the realization of the
periodic time-dependent potential can be achieved by modulating
gate voltages applied to the structure.
In recent years parametric pumping has attracted a considerable
attention within the emerging field of
spintronics\cite{wolf_spintronics}, as a technique to generate
spin-polarized currents in semiconductors\cite{datta,chao_sp}.
This widens the field of usual magneto-electronics in metals and
opens the possibility of combining the rich physics of
spin-polarized particles with all the advantages of semiconductor
fabrication and technology. Besides, since spin relaxation times
of semiconductors can be rather long and coherence of spin states
can be maintained up to scales more than 100$\mu$m \cite{kikkawa},
an intriguing question is whether a pure spin current can be
pumped in absence of a charge current. This is the case when equal
amounts of electrons with opposite spins move in opposite
direction in the system. The possibility of generating pure spin
currents is attracting enormous interest, not only as a
theoretical study {\it per se}, but also in view of proposed
future applications, including scalable devices for quantum
information processes\cite{quantum_information}. Several works
focused on a variety of spin pumps have recently
appeared\cite{brataas_sp,mucciolo_sp,watson_sp,general_sp,aono_sp}.

In this Letter, we report  results on quantum pumping through a
one-dimensional ring shaped conductor interrupted by a tunnel
barrier in presence of time-reversal Aharonov-Casher effect and
Aharonov-Bohm magnetic flux and analyze a new scheme of realizing
pure spin pumping in mesoscopic systems. An important feature of
ring shaped conductors is the appearance of quantum interference
effects under the influence of electromagnetic potentials, known
as Aharonov-Bohm\cite{aharonov_bohm} and
Aharonov-Casher\cite{aharonov_casher} effect.  The Aharonov-Bohm
(AB) phase acquired by a charged particle encircling a magnetic
flux is an example of topological phase known for a long time. In
1984, Aharonov and Casher noticed the dual of the AB effect known
as AC effect. The AC effect originates from the spin-orbit (SO)
coupling between the moving magnetic dipole and the electric
field. It is expected to implement the spin current modulation
manifested by the AB effect. The AC effect in AB rings with Rashba
spin-orbit coupling has been recently investigated both
theoretically and experimentally\cite{ring_exp}.  In numerous
studies, the transmission properties of mesoscopic AB and AC rings
coupled to current leads were studied under various aspects such
as AB flux and coupling dependence of
resonances\cite{buttiker_ring,sigrist_as}, geometric (Berry)
phases\cite{loss_ring,stern_ring,aronov_ring,qian_ring,hentschel_ring}
and spin flip, precession, interference
effects\cite{yi_ring,choi_ring,nitta_ring,frustaglia_ring,molnar_ring_ab_ac,capozza_ring}.
Persistent currents in absence of current leads were studied in
Ref.[\onlinecite{balatsky_ring}]. In our proposal we consider the
one-dimensional ring with a time-dependent tunnel barrier and a
time oscillating radial (or transverse) electric field in absence
of external bias and shall show that dc spin and charge currents
are induced by parametric pumping procedure. We calculate
analytically the spin dependent transmission and reflection
coefficients and apply the scattering matrix
approach\cite{scattering_theory} to calculate the current. Due to
the interplay of spin reversal symmetry breaking due to the AC
effect and of time reversal symmetry breaking intrinsic in the
pumping procedure, a pure spin current can be generated and we
shall analyze its behavior in the weak and strong pumping regime.

\begin{figure}[t]
\centering
\includegraphics[width=7cm]{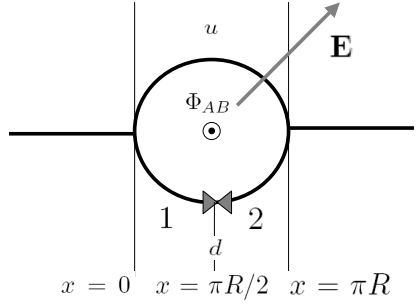}\\
\caption{Aharonov-Bohm-Casher ring.  A delta barrier potential
($\triangleright\triangleleft$) is placed in the lower branch of
the ring.} \label{fig:ring}
\end{figure}

We consider noninteracting electrons confined to a one-dimensional
ring of a radius $R$ with two leads embedded in a radial electric
field $\mathbf{E}$. A magnetic flux is  assumed to be set along
the axis of the same ring. We assume that the length of the ring
is smaller than the spin diffusion length, so to neglect spin-flip
processes. The ring is interrupted by a single tunnel barrier in
one half of the ring, enabling  to apply a well defined potential.
The one-particle Hamiltonian  of our system is given by
$\hat{H}=\hat{H}_R+\hat{V}$, where
\begin{equation}
\hat{H}_R=\frac{1}{2m^\star}(\hat{\mathbf{P}}-\frac{\mu}{2c}
\hat{\mathbf{\sigma}} \times \mathbf{E}-\frac{e}{c} \mathbf{A})^2,
\end{equation}
$\hat{\mathbf{P}}$ is the momentum of the electron, $c$ is the
velocity of light, $\mu=g \mu_B$ with $\mu_B$ being the Bohr
magneton, $g$ is the spin g-factor and $m^\star$ is the effective
mass of the carriers. The linear term $\hat{\mathbf{\sigma}}\cdot
(\mathbf{E}\times \hat{\mathbf{P}})$ represents the spin-orbit
coupling, $\sigma_i (i=1,2,3)$ are the Pauli matrices;
$\mathbf{A}=\frac{\Phi_{AB}}{2\pi R} \mathbf{e}_\phi$ is the
vector potential field of the magnetic flux $\Phi_{AB}$ with zero
magnetic field. ${\hat V}$ is the tunnel barrier potential that we
model by a delta function in lower branch of the ring (see Fig.1).
To start with, we consider a radial electric field $\mathbf{E}=E
\mathbf{e}_r$, with $\mathbf{e}_r$ denoting the unit vector along
the radial direction, and put $V=0$. In the case of a
one-dimensional ring, a confining potential $U(r)$ needs to be
added in order to force the electron wave function to be localized
on the ring. A simple possibility is the harmonic potential
$U(r)=\frac{1}{2} K (r-R)^2$, where $R$ is the radius of the ring.
Considering only the lowest radial
mode\cite{meijer_ham_ring,zhou_ham_ring}, the resulting effective
1D Hamiltonian for fixed radius $R$ is:
\begin{equation}
\label{eq:ham} \hat{H}_R^{1D}=\frac{\hbar^2}{2m^\star R^2}(-i
\frac{\partial}{\partial\phi}-\frac{\mu E R}{2 \hbar c} \sigma_z
-\frac{e \Phi_{AB}}{ h c})^2.
\end{equation}
The eigenfunctions of the Hamiltonian are given by
$\Psi_{n,\sigma}=1/\sqrt{2\pi}e^{in \phi}\chi_\sigma$ where
$\sigma=\pm$ denotes the spin up and down along the $z$ direction
with $\chi_+\equiv(1,0)^T, \mbox{ }\chi_-\equiv (0,1)^T$, $T$
denotes the transpose of the matrix. The corresponding eigenvalues
are $E_{n,\sigma}=\hbar^2/(2m^\star R^2) (n-\Phi_\sigma/2\pi)^2$
and $n$ is an integer. The phase $\Phi_\sigma=\Phi_{AB}+\sigma
\Phi_{AC}$ is the total phase that the electron acquires while the
two spin states $\Psi_{n,\sigma}$ move in the ring in the presence
of the electric field ($\Phi_{AC}=\pi \theta_0$,
$\theta_0=\frac{\mu E R}{\hbar c}$) and of the magnetic flux
($\Phi_{AB}$)\cite{frustaglia_ring}. It is instructive to estimate
the magnitude of the AC flux in realistic mesoscopic systems. For
a ring of radius $10^{-6}$m , in external electric field $E\sim
10^{4}$V/cm, we have $\Phi_{AC}/\Phi_0\sim 10^{-3}$ (where
$\Phi_0=hc/e$ is the quantum of flux) for particle with
gyromagnetic ratio $g\sim 1$. On the other hand, in semiconductors
$g$ can be two orders of magnitude larger and the AC flux becomes
of the order $10^{-1}$ which makes the interference effect
associated with the AC effect experimentally observable. Assumed
that electrons in the two leads are free and have momentum $k$,
the corresponding energy is $\hbar^2 k^2/2m^\star$. When an
electron moves along the upper arm in the clockwise direction from
the input intersection at $x=0$ (see Fig.\ref{fig:ring}), it
acquires a phase $\Phi_\sigma/2$ at the output intersection $x=\pi
R$, whereas the electron acquires a phase $-\Phi_\sigma/4$ in the
counterclockwise direction along the other arm when moving from
$x=0$ to $x=\pi R/2$ and from $x=\pi R/2$ to $x=\pi R$,
respectively. Therefore the total phase is $\Phi_\sigma$ when the
electron goes through the loop. The electric field in the ring
changes the momenta of the electrons in different spin states
$\chi_\pm$ as $k_c^\sigma=k+\Phi_\sigma/2\pi R$ and
$k_a^\sigma=k-\Phi_\sigma/2\pi R$, where the subscripts denote the
clockwise and counterclockwise direction. The wave functions in
the upper($u$) and lower($d$) arm of the ring can be written as:
\begin{eqnarray}
\Psi_{u}=\sum_{\sigma=\pm} (c_{u,\sigma}e^{ikx}+d_{u,\sigma}e^{-ikx})e^{i\Phi_\sigma x/2 \pi R}\chi_\sigma, \nonumber \\
\Psi_{d\alpha}=\sum_{\sigma=\pm}
(c_{d\alpha,\sigma}e^{ikx}+d_{d\alpha,\sigma}e^{-ikx})e^{-i\Phi_\sigma
x/2 \pi R}\chi_\sigma,
\end{eqnarray}
where $x$ ranges from $0$ to $\pi R$ and we have assumed that the
electron travels in the $x$ direction. The index $d\alpha=1,2$
denotes the wave function in the two-halves of the lower branch.
The wave function of the electron incident from the left lead in
the left and right electrodes is:
\begin{equation}
\psi_L=\psi_i+(r_\uparrow,r_\downarrow)^Te^{-ik x},\mbox{
}\psi_R=(t_\uparrow,t_\downarrow)^T e^{ik x},
\end{equation}
where $r_\sigma$ and $t_\sigma$ are the spin-dependent reflection
and transmission coefficients, $\psi_i$ is the wave function of
the injected electron $\psi_{i}=e^{ik x}\chi_\sigma$. For an
incident electron from the right lead an analogous expansion in
terms of reflection and transmission coefficients is possible with
${i,r_\sigma}$ (for left lead) and ${t_\sigma,0}$ (for right lead)
replaced by ${0,t'_\sigma}$ and ${r'_\sigma,i'}$. This enables us
to formulate the scattering matrix equation of the ring system as
$\hat{O}=\hat{S}\hat{I}$, where $\hat{O},\hat{I}$ stand for
outgoing and incoming wave coefficients. In the parametric pumping
theory, the quantum mechanical current pumped in the ring is
related to the derivative of the scattering matrix elements with
respect to time-dependent parameters. To calculate the scattering
matrix elements we use the local coordinate
system\cite{xia_local_coord}. The Griffith boundary
conditions\cite{griffith_boundary} state that the wave function is
continuous and that the current density is conserved at each
intersection. At the point $x=\pi R/2$ in the lower branch the
same conditions apply in presence of the delta tunnel barrier.
After some algebra we obtain the transmission coefficients
$t_\sigma(kR,\Phi_\sigma,z)$ where $z=2m^\star V/k\hbar^2$ and $V$
is the amplitude of the delta barrier. The explicit expression of
$t_\sigma(kR,\Phi_\sigma,z)$ is:
\begin{eqnarray}
\frac{8\,\sin (\frac{\pi k R}{2})\left( -4\,\cos (\frac{\pi k
R}{2})\,\cos (\frac{\Phi_{\sigma}}{2} ) + z\,\sin (\frac{\pi k
R}{2})\,e^{i\frac{\Phi_{\sigma}}{2}}\right)}
    {4\,z\,\cos (\pi k R) - 2\,\left( 5\,i + 2\,z \right) \,\cos (2\pi k R) +
    i\,\left( 2 + 8\,\cos (\Phi_{\sigma} ) - 2\,z\,\sin (\pi k R) + \left( 8\,i + 5\,z \right) \,\sin (2\pi k R) \right)}.
\end{eqnarray}
In the limit $z\rightarrow 0$, $t_\sigma(
kR,\Phi_\sigma,z\rightarrow 0)= i\cos( \pi k R)\sin
(\Phi_\sigma/2)/\lbrack \sin^2(\Phi_\sigma/2)-(\cos(\pi k R)-i/2
\sin (\pi k R))^2 \rbrack$.
The corresponding transmission probability is
$T=|t_\uparrow|^2+|t_\downarrow|^2$. Our general expression shows
that $T$ is independent of the incident spin state. Both the
expression of the transmission amplitude and probability are
determined by the tunnel barrier strength, the total phase, the
kinetic state of the incident electrons, the electric field and
the magnetic flux. In the framework of the Landauer-Buttiker
theory\cite{landauer_cond}, the quantum mechanical transmission
amplitude is related to the conductance.
In order to generate a pure spin current in our system, we propose
the use of adiabatic quantum
pumping\cite{brouwer,scattering_theory}. To inject a spin or
charge current in the lead 1 one has to modulate two independent
(out of phase) parameters of the device in absence of external
bias. Adiabatic modulation procedure of the out-of-phase pumping
amplitudes dynamically breaks time reversal invariance which in
turn leads to a net spin current being pumped. Namely, we
adiabatically modulate the Aharonov-Casher flux (i.e. the electric
field or the spin-orbit coupling) and the strength of the contact
barrier in the lower branch of the ring as:
$\Phi_{AC}=\Phi^{0}_{AC}+\Phi^{\omega}_{AC}\sin(\omega t+\varphi),
\mbox{} z=z_{0}+z_{\omega}\sin(\omega t)$, where $\varphi$ is the
phase difference. In the weak pumping regime, the injected current
in a given lead is proportional to the area enclosed in the
parameters space, $A_0={\sin(\varphi)\over 2
\pi}z_{\omega}\Phi^{\omega}_{AC}$ . The $\sin \varphi$ behavior is
lost in the strong pumping regime ($\Phi^{0}_{AC}\ll
\Phi^{\omega}_{AC}$, $z_{0}\ll z_{\omega}$). In the zero
temperature limit, the current pumped with arbitrary spin $\sigma$
in the lead 1 can be derived by the following formula
\cite{brouwer}:
\begin{eqnarray}
I_{1\sigma}={\omega e\over 2 \pi}\int_{0}^{\tau}dt {\sum_{l=1,2}{d
N_{1\sigma}\over d X_{l}}{d X_{l}\over d t}},
\end{eqnarray}
wherein $\tau={2\pi\over\omega}$ is the period of the forcing
signals, $\omega$ is the pumping frequency and $e$ represents the
electron charge. The quantity,  ${d N_{1\sigma}\over d X_{l}}$ is
the so-called electronic
injectivity\cite{buttiker_injectivity,wang2003} and is given by:

\begin{eqnarray}
{d N_{1\sigma}\over d X_{l}}={1\over 2 \pi}\Im
\{{\sum_{j=1,2}\mathcal{S}^{\sigma \ast
}_{1j}\partial_{X_{l}}\mathcal{S}_{1j}^{\sigma}}\},
\end{eqnarray}
with $l=1,2$. The pumping parameters are $X_{1}\doteq z$ and
$X_{2}\doteq \Phi_{AC}$, $\Im$ denotes the imaginary part, while
$j=1,2$ denotes the left lead or right lead. Therefore, the charge
and spin currents, $I_{ch}$ and $I_{sp}$, in the first lead are:
\begin{eqnarray}
I_{ch}=I_{1\uparrow}+I_{1\downarrow}, \mbox{ }
I_{sp}=I_{1\uparrow}-I_{1\downarrow}.
\end{eqnarray}

\begin{figure}[htbp]
\centering
\includegraphics[width=15cm]{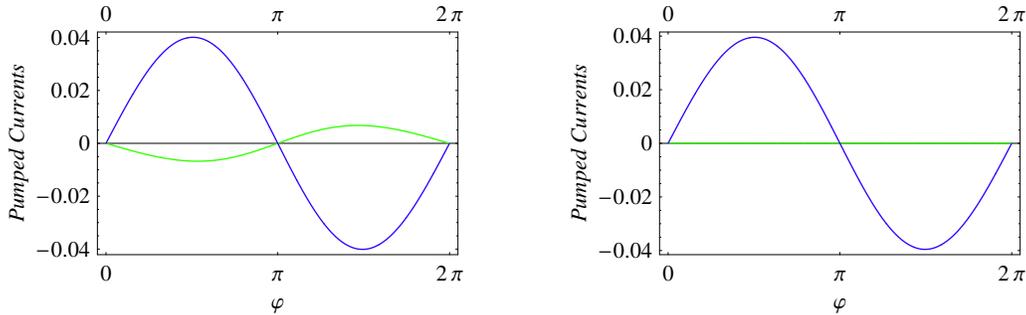}\\
\caption{(Color online) Pumped charge current, green(bright gray)
line, and spin current, blue(dark gray) line, in units of ${\omega
e \over 2 \pi}$ as a function of the phase shift $\varphi$, in the
weak pumping regime. The model parameters in the left panel are:
$k R=10$, $\Phi _{AB}=0.45$, $\Phi^{0}_{AC}=0.1$,
$\Phi^{\omega}_{AC}=0.1$, $z_0=0.1$ and $z_{\omega}=0.1$. In the
right panel a pure spin current is obtained for $\Phi_{AB}=0.5$.
Fluxes are in units of $2\pi$.} \label{fig:currents_wp}
\end{figure}

\begin{figure}[htbp]
\centering
\includegraphics[width=7cm]{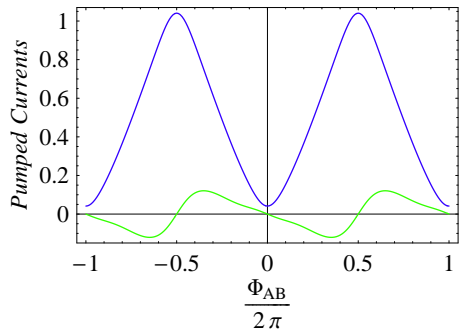}\\
\caption{(Color online) Pumped charge current, green(bright gray)
line, and spin current, blue(dark gray) line, in units of ${\omega
e \over 2 \pi}$ as a function of the Aharonov-Bohm flux
$\Phi_{AB}$. Parameters: $k R=10$, $\varphi=2.6$,
$\Phi^{0}_{AC}=-0.15$, $\Phi^{\omega}_{AC}=0.4$, $z_0=1$ and
$z_{\omega}=1.8$.} \label{fig:currents_ab}
\end{figure}

The dc charge and spin currents generated by the pumping mechanism
are shown in Fig.\ref{fig:currents_wp} as a function of $\varphi$
in the weak pumping regime for two different values of the AB
flux. As shown a {\it pure} spin current is generated when the AB
flux is half-integer. We also find that in the weak pumping
regime, the charge and spin pumped per cycle are not quantized. In
the strong pumping regime, a pure spin current can be obtained
once we properly tune the phase $\varphi$. For specific values of
$\varphi$, we find that the pumped spin and charge current
oscillate with the AB flux and a pure spin current is obtained at
half-integer values of the AB flux, as shown in
Fig.\ref{fig:currents_ab}.

\begin{figure}[htbp]
\centering
\includegraphics[width=7cm]{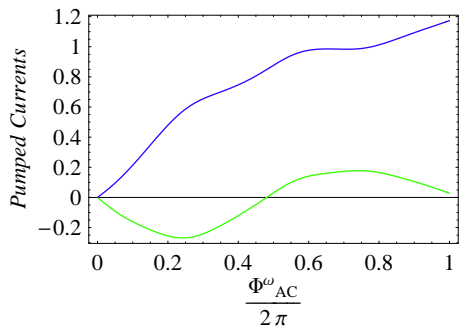}\\
\caption{(Color online) Pumped charge current, green(bright gray)
line, and spin current, blue(dark gray) line, in units of ${\omega
e \over 2 \pi}$ as a function of the pumping parameter
$\Phi_{AC}^\omega$ for the same parameters of
Fig.\ref{fig:currents_ab}. } \label{fig:currents_ac}
\end{figure}

From the analytical expression of the pumped currents in the weak
pumping regime, we find that $I_\sigma$ is proportional to
$\frac{\sigma}{2} A_0 \sin (\Phi_{AB}+\sigma \Phi_{AC})$.  Thus,
the spin current is maximum at half-integer values of the AB flux
(in units of $2\pi$) or at integer values of the AC flux, while
the charge current is zero. In Fig.\ref{fig:currents_ac} the
pumped spin and charge currents are shown as a function of  the
pumping amplitude $\Phi_{AC}^\omega$ from weak to strong pumping
regime. We find a finite spin current with a small charge current.
The figure also conveys the very important fact that for the
entire range the magnitude of pumped spin current increases, which
suggests that our model device would pump large spin currents in
the very strong pumping regime. Further the pumped charge current
is, throughout the range of the pumping amplitude, constant in
average. The present analysis also show that $\Phi_{AC}^\omega$
can be tuned to a regime where the spin transport through the
system is quantized, i.e. the average number of spins transmitted
in a cycle is integer. This feature allows us to consider the
proposed pumping system as a noiseless pump device working in an
optimal way.

Since the AC phase can be mapped to the spin-orbit interaction
parameter, our results extend to Rashba-gate-controlled
rings\cite{ring_exp}. In this case, the Rashba electric field
results from asymmetric confinement along the z direction
perpendicular to the plane of the ring that can be properly tuned
by a gate voltage. An effective Hamiltonian analogous to
(\ref{eq:ham}) is derived\cite{meijer_ham_ring}, with $-\frac{\mu
E R}{ \hbar c} \sigma_z$ replaced by $\frac{\alpha m^\star R}{2
\hbar^2}\sigma_r$\cite{molnar_ring_ab_ac}, where $\alpha$
represents the Rashba electric field along $z$ and is a tunable
quantity. In this case, the AC phase is given by: $\Phi_{AC}=-\pi
\sqrt{1+2\alpha \frac{m^\star R}{\hbar^2}}$. A complete derivation
of eigenvalues and eigenfunctions can be found in
Ref.[\onlinecite{molnar_ring_ab_ac}], and our expression of
transmission and reflection amplitudes is essentially unchanged.
For an InGaAs based two-dimensional electron gas, $\alpha$ can be
controlled by a gate voltage with typical values in the range
(0.5-2.0)$\times 10^{-11} $eVm, which in turn will correspond to
values of the AC flux between 0.3$\pi$ and 3$\pi$ for the
effective mass of InAs $m^\star=0.023 m_e$ and radius $R=0.25
\mu$m. In conclusion, we have shown a novel scheme of parametric
pumping of charge and spin in mesoscopic system. The generation of
a pure spin current in AB-AC ring results from the interference
effect of electrons with different topological phases and from the
adiabatic modulation of two out-of-phase pumping amplitudes that
dynamically breaks the time reversal symmetry.  By changing the
applied electric field or the SO coupling, the AC phase
contribution to the current oscillation can be tuned to give a
pure spin current thanks to spin reversal symmetry breaking. Our
proposal is within reach with today's technology for experiments
in  semiconductor heterostructures with a two dimensional electron
gas (2DEG) which has an internal electric field due to an
asymmetric quantum well\cite{ring_exp_AC} and the spin current in
our device could be measurable by the scanning capacitance probe
technique. Indeed, spin interference effects in
Rashba-gate-controlled ring with a quantum point contact inserted
have recently been reported\cite{ring_exp}.

{\it Acknowledgments} The authors acknowledge enlightening
discussions with M. Marinaro and S. Cojocaru and E. Orignac for
comments on the manuscript.

\bibliographystyle{prsty}
\bibliography{pumpring}

\end{document}